\documentclass{article}
\usepackage{amssymb}
\usepackage{amsmath}
\usepackage{hyperref}

\input{tcilatex}
\begin{document}

\title{Massive Dual Spin 2 Revisited}
\author{T.L. Curtright$^{\S \dag }$\ and H. Alshal$^{\S \bigtriangleup \ast
} $ \\
$^{\S }$Department of Physics, University of Miami, Coral Gables, Florida
33124\\
$^{\bigtriangleup }$Department of Physics, Faculty of Science, Cairo
University, Giza, 12613, Egypt\\
$^{\dag }${\footnotesize curtright@miami.edu\ \ \ \ \ }$^{\ast }$%
{\footnotesize halshal@sci.cu.edu.eg}}
\date{}
\maketitle

\begin{abstract}
We reconsider a massive dual spin 2 field theory in four spacetime
dimensions. \ We obtain the Lagrangian that describes the lowest order
coupling of the field to the four-dimensional curl of its own
energy-momentum tensor. \ We then find some static solutions for the dual
field produced by other energy-momentum sources and we compare these to
similar static solutions for non-dual \textquotedblleft finite
range\textquotedblright\ gravity. \ Finally, through use of a nonlinear
field redefinition, we show the theory is the exact dual of the
Ogievetsky-Polubarinov model for a massive spin 2 field.

\noindent \hrulefill
\end{abstract}

\bigskip

\begin{center}
In tribute to \href{https://en.wikipedia.org/wiki/Peter_Freund}{Peter George
Oliver Freund} (1936-2018)
\end{center}

\bigskip

\section{Introduction}

We reconsider research first pursued\ long ago in collaboration with Peter
Freund at the University of Chicago \cite{CF1980}, in a continuing effort to
tie up some loose ends. \ \textit{Inde est, quod.}

In the winter of 1979-1980, while thinking about strings\footnote{%
The set of ghost fields required for the covariant quantization of
antisymmetric gauge fields, as described in the letter to Paul Townsend
linked \href{http://www.physics.miami.edu/~curtright/GHOSTS1979.GIF}{here},
was later generalized to arbitrary tensor gauge fields, as described in 
\href{http://inspirehep.net/record/234762/files/CountingStringStates.pdf}{a
conference contribution} (see Section V of \cite{C1986}).} as a
post-doctoral fellow in Yoichiro Nambu's theory group at The Enrico Fermi
Institute, one of us (TLC) proposed gauge theories for massless fields that
were neither totally symmetric nor totally antisymmetric Lorentz
representations, and also discussed the effects of adding mass terms for
such fields \cite{C1980}. \ Peter found this\ to be very interesting and
worth further consideration, perhaps because of his previous work on massive
scalars \cite{FN1968} and finite range gravity \cite{FMS1968}.

The \href{https://en.wikipedia.org/wiki/Curtright_field}{simplest example}
is $T_{\left[ \lambda \mu \right] \nu }$ with permutation symmetries $T_{%
\left[ \lambda \mu \right] \nu }=-T_{\left[ \mu \lambda \right] \nu }$ and $%
T_{\left[ \lambda \mu \right] \nu }+T_{\left[ \mu \nu \right] \lambda }+T_{%
\left[ \nu \lambda \right] \mu }=0$. \ While the theory has remarkable
differences between the massless and massive cases in any number of
spacetime dimensions, in 4D Minkowski space the model has a particularly
striking mass discontinuity, with \emph{no} propagating modes when massless
but with five modes corresponding to angular momentum $J=2\hbar $ when
massive \cite{C1980}.

A consistent field equation for the massive 4D model was subsequently
proposed in \cite{CF1980}, namely,%
\begin{equation}
\left( \square +m^{2}\right) T_{\left[ \lambda \mu \right] \nu }=\kappa
\left( 2\varepsilon _{\lambda \mu \alpha \beta }\partial ^{\alpha }\Theta
_{\ \nu }^{\beta }+\varepsilon _{\nu \mu \alpha \beta }\partial ^{\alpha
}\Theta _{\ \lambda }^{\beta }-\varepsilon _{\nu \lambda \alpha \beta
}\partial ^{\alpha }\Theta _{\ \mu }^{\beta }\right) ~,  \label{FE}
\end{equation}%
where $\Theta _{\mu \nu }$ is any conserved, symmetric tensor, e.g. the
energy-momentum tensor, and $\kappa $ is a dimensionful parameter with units 
$1/m^{2}$, most naturally related to Newton's constant and the Hubble length.%
\footnote{%
See \cite{CF1980}\ p 418$.$} \ This field equation implies that the trace
and all divergences of $T_{\left[ \lambda \mu \right] \nu }$ decouple, i.e.
they\ are free fields, and therefore they may be consistently set to zero
leaving a model with unadulterated spin 2 particles of mass $m$. \ 

That said, \cite{CF1980}\ stopped short and did \emph{not} provide a
Lagrangian that led to the field equation (\ref{FE}) for the case where $%
\Theta _{\mu \nu }$ depends only on $T_{\left[ \lambda \mu \right] \nu }$\
itself. \ This shortcoming is remedied here, albeit only to lowest order in $%
\kappa $. \ However, there seem to be no fundamental principles to prevent
extension of the result to all orders in $\kappa $, thereby emulating the
theory of a self-coupled dual scalar as discussed in\ a companion paper \cite%
{C2019}.\newpage

\section*{Properties of the massless\ free field theory}

There are three increasingly compact forms for the \emph{massless} $T_{\left[
\lambda \mu \right] \nu }$ \emph{free field} Lagrangian density, involving
three terms, two terms, and one term, respectively.%
\begin{eqnarray}
\mathcal{L} &=&-\frac{1}{4}\left( R_{\left[ \lambda \mu \right] \left[ \nu
\rho \right] }R^{\left[ \lambda \mu \right] \left[ \nu \rho \right]
}-4R_{\lambda \nu }R^{\lambda \nu }+R^{2}\right)  \notag \\
&=&-\frac{1}{6}\left( F_{\left[ \lambda \mu \nu \right] \rho }~F^{\left[
\lambda \mu \nu \right] \rho }-3F_{\left[ \mu \nu \right] }~F^{\left[ \mu
\nu \right] }\right)  \notag \\
&=&+\frac{1}{36}~K_{\mu \nu }K^{\nu \mu }\ .  \label{Simplest}
\end{eqnarray}%
The first two forms are legitimate for any number of spacetime dimensions 
\cite{C1980} but the last is specific to 4D. \ This is evident in the \emph{%
flat} spacetime definitions%
\begin{eqnarray}
R_{\left[ \lambda \mu \right] \left[ \nu \rho \right] } &=&\partial _{\nu
}T_{\left[ \lambda \mu \right] \rho }-\partial _{\rho }T_{\left[ \lambda \mu %
\right] \nu }\ ,  \notag \\
R_{\lambda \nu } &=&g^{\mu \rho }R_{\left[ \lambda \mu \right] \left[ \nu
\rho \right] }\ ,  \notag \\
R &=&g^{\lambda \nu }R_{\lambda \nu }\ .
\end{eqnarray}%
\begin{gather}
F_{\left[ \lambda \mu \nu \right] \rho }=\partial _{\lambda }T_{\left[ \mu
\nu \right] \rho }+\partial _{\mu }T_{\left[ \nu \lambda \right] \rho
}+\partial _{\nu }T_{\left[ \lambda \mu \right] \rho }\ ,  \notag \\
F_{\left[ \mu \nu \right] }=g^{\lambda \rho }F_{\left[ \lambda \mu \nu %
\right] \rho }\ .  \label{FStrength}
\end{gather}%
and finally, for flat 4D spacetime,\footnote{%
Moreover, in 4D spacetime%
\begin{equation*}
R_{\left[ \lambda \mu \right] \left[ \nu \rho \right] }R^{\left[ \lambda \mu %
\right] \left[ \nu \rho \right] }-4R_{\lambda \nu }R^{\lambda \nu }+R^{2}=-%
\frac{1}{4}~\varepsilon ^{\lambda \mu \gamma \delta }~\varepsilon ^{\alpha
\beta \nu \rho }~R_{\left[ \lambda \mu \right] \left[ \nu \rho \right] }~R_{%
\left[ \alpha \beta \right] \left[ \gamma \delta \right] }
\end{equation*}%
If $R_{\left[ \lambda \mu \right] \left[ \nu \rho \right] }$ were the
Riemann curvature, such that $R_{\left[ \lambda \mu \right] \left[ \nu \rho %
\right] }=R_{\left[ \nu \rho \right] \left[ \lambda \mu \right] }$, this
would be the Euler density. \ But as defined here, $R_{\left[ \lambda \mu %
\right] \left[ \nu \rho \right] }\neq R_{\left[ \nu \rho \right] \left[
\lambda \mu \right] }$, so $\mathcal{L}$\ is \emph{not} a total divergence
and its action is \emph{not} just topological.}%
\begin{equation}
K_{\mu }^{\ \nu }=F_{\left[ \alpha \beta \gamma \right] \mu }\varepsilon
^{\alpha \beta \gamma \nu }=3\left( \partial _{\alpha }T_{\left[ \beta
\gamma \right] \mu }\right) \varepsilon ^{\alpha \beta \gamma \nu }\ .
\label{KDefn}
\end{equation}%
This last definition immediately leads to the kinematic identities 
\begin{equation}
K_{\mu }^{\ \mu }\equiv 0\ ,\ \ \ \partial ^{\nu }K_{\mu \nu }\equiv 0\ ,
\label{Trace&DivConditions}
\end{equation}%
where the first of these is a consequence of $T_{\left[ \lambda \mu \right]
\nu }+T_{\left[ \mu \nu \right] \lambda }+T_{\left[ \nu \lambda \right] \mu
}=0$.

Gauge invariance of the theory is most easily seen in 4D for the bilinear-in-%
$K$ form of $\mathcal{L}$. \ A gauge transformation on the $T_{\left[
\lambda \mu \right] \nu }$ field is \cite{C1980}%
\begin{equation}
\delta T_{\left[ \lambda \mu \right] \nu }=\partial _{\lambda }S_{\mu \nu
}-\partial _{\mu }S_{\lambda \nu }+\partial _{\lambda }A_{\mu \nu }-\partial
_{\mu }A_{\lambda \nu }+2\partial _{\nu }A_{\mu \lambda }\ ,
\end{equation}%
where $S_{\mu \nu }$ is any local, differentiable, symmetric tensor field,
and $A_{\mu \nu }$ is any local, differentiable, antisymmetric tensor field.
\ This in turn leads to the gauge transformation of the $F_{\left[ \alpha
\beta \gamma \right] \mu }$ field strength \cite{C1980}%
\begin{equation}
\delta F_{\left[ \alpha \beta \gamma \right] \mu }=-2\partial _{\mu }\left(
\partial _{\alpha }A_{\beta \gamma }+\partial _{\beta }A_{\gamma \alpha
}+\partial _{\gamma }A_{\alpha \beta }\right) \ ,
\end{equation}%
and thence to the gauge transformation of $K_{\mu \nu }$, but only in 4D,%
\begin{equation}
\delta K_{\mu \nu }=\partial _{\mu }\omega _{\nu }\ ,  \label{KGaugeTrans}
\end{equation}%
where $\omega _{\nu }=-6\left( \partial ^{\alpha }A^{\beta \gamma }\right)
\varepsilon _{\alpha \beta \gamma \nu }$ is a differentiable, \emph{%
divergenceless} pseudo-vector, but otherwise arbitrary. \ This form of $%
\delta K_{\mu \nu }$ manifestly maintains the trace and divergence
conditions for $K_{\mu \nu }$ as given in (\ref{Trace&DivConditions}). \ 

Gauge invariance of the action in 4D, modulo surface contributions, then
follows immediately from the divergence condition in (\ref%
{Trace&DivConditions}). \ For any 4-volume $V_{4}$,%
\begin{equation}
\delta \int_{V_{4}}K_{\mu \nu }K^{\nu \mu }d^{4}x=2\int_{V_{4}}K_{\mu \nu
}\delta K^{\nu \mu }d^{4}x=2\int_{V_{4}}K_{\mu \nu }\partial ^{\nu }\omega
^{\mu }d^{4}x=2\int_{\partial V_{4}}\omega ^{\mu }K_{\mu \nu }n^{\nu
}d^{3}x\ ,
\end{equation}%
where $n^{\nu }$ is the local normal vector on the boundary $\partial V_{4}$%
. \ For gauge transformations that vanish on $\partial V_{4}$ the action is
therefore invariant.

\subsubsection*{There are no physical states for the massless free field
theory in 4D}

The bilinear-in-$K$ form for the action is the most transparent way to see
the only solutions of the free field equations in 4D are just gauge
transformations. \ With $K_{\mu \nu }$ defined by (\ref{KDefn}), the bulk
field equations that follow from $\int K_{\mu \nu }K^{\nu \mu }d^{4}x$ are
simply 
\begin{equation}
\varepsilon _{\alpha \beta \lambda \mu }\partial ^{\lambda }K^{\mu \nu }=0\ ,
\end{equation}%
along with the conditions (\ref{Trace&DivConditions}). \ That is to say, 
\begin{equation}
\partial ^{\lambda }K^{\mu \nu }-\partial ^{\mu }K^{\lambda \nu }=0\ ,
\end{equation}%
and therefore $K_{\mu \nu }$ must have the form (\ref{KGaugeTrans}). \ So
on-shell $K_{\mu \nu }$ is just a gauge transformation. \ Thus there are no
physical degrees of freedom for the massless, free $T_{\left[ \lambda \mu %
\right] \nu }$ field in 4D.

Moreover, on-shell conservation of energy-momentum for the flat space 4D
theory is as easy as one might expect for fields which are local gauge
transformations. \ Consider%
\begin{equation}
\vartheta _{\mu }^{\ \nu }=K_{\mu \beta }K^{\beta \nu }-\frac{1}{2}~\delta
_{\mu }^{\ \nu }K_{\alpha \beta }K^{\beta \alpha }\ .  \label{NonSymEMTensor}
\end{equation}%
On-shell in 4D $K_{\alpha \beta }=\partial _{\alpha }\omega _{\beta }$ with $%
\partial ^{\mu }\omega _{\mu }=0$, hence%
\begin{equation}
\partial _{\nu }\left( K_{\mu \beta }K^{\beta \nu }\right) =\left( \partial
_{\nu }\partial _{\mu }\omega _{\beta }\right) \partial ^{\beta }\omega
^{\nu }=\left( \partial _{\mu }\partial _{\nu }\omega _{\beta }\right)
\partial ^{\beta }\omega ^{\nu }=\frac{1}{2}~\partial _{\mu }\left(
K_{\alpha \beta }K^{\beta \alpha }\right) \ ,
\end{equation}%
and therefore the tensor (\ref{NonSymEMTensor}) is conserved on-shell, $%
\partial _{\nu }\vartheta _{\mu }^{\ \nu }=0$. \ Indeed, $\vartheta _{\mu
}^{\ \nu }$ encodes the physics of the massless $T_{\left[ \lambda \mu %
\right] \nu }$ model in 4D because both the energy and spatial momentum
densities are given on-shell by spatial derivatives, namely, $\vartheta
_{00}=\overrightarrow{\nabla }\cdot \left( -\frac{1}{2}\left( 
\overrightarrow{\omega }\cdot \overrightarrow{\nabla }\right) 
\overrightarrow{\omega }\right) $ and $\vartheta _{k0}=\partial _{k}\left(
\omega _{0}\partial _{0}\omega _{0}\right) +\overrightarrow{\nabla }\cdot
\left( \omega _{0}\partial _{k}\overrightarrow{\omega }\right) $. \
Consequently, fields that vanish sufficiently rapidly as $r\rightarrow
\infty $ (i.e. $\lim_{r\rightarrow \infty }r\omega ^{\alpha }=0$) carry
neither net energy nor net momentum, as would be expected for configurations
that are just gauge transformations.

Also, by definition the tensor $\vartheta _{\mu \nu }$ is not manifestly
symmetric because in general $K_{\mu \nu }\neq K_{\nu \mu }$. \ Rather,%
\begin{equation}
K_{\mu \nu }=K_{\nu \mu }+3\varepsilon _{\mu \nu \alpha \beta }F_{\ \ \ \ \
\ \lambda }^{\left[ \alpha \beta \lambda \right] }\ .
\end{equation}%
But this is remedied for the massless theory by imposing the Lorentz
covariant gauge condition%
\begin{equation}
F_{\ \ \ \ \ \ \lambda }^{\left[ \alpha \beta \lambda \right] }=\partial
^{\lambda }T_{\ \ \ \ \lambda }^{\left[ \alpha \beta \right] }+\partial
^{\alpha }T_{\ \ \ \ \lambda }^{\left[ \beta \lambda \right] }-\partial
^{\beta }T_{\ \ \ \ \lambda }^{\left[ \alpha \lambda \right] }=0\ ,
\end{equation}%
a condition preserved by gauge transformations so long as $\omega _{\alpha
}=\partial _{\alpha }\psi $ with $\square \psi =0$. \ In this class of
gauges, both $K_{\mu \nu }=K_{\nu \mu }$ and $\vartheta _{\mu \nu
}=\vartheta _{\nu \mu }$. \ 

Moreover, in this class of gauges there are some obvious similarities with
Galileon theory \cite{CF2012} in that the $K$-tensor is a Hessian matrix, 
\begin{equation}
K_{\mu \nu }=\partial _{\mu }\partial _{\nu }\psi \ ,
\end{equation}%
albeit traceless. \ This may then be combined with the expansion of a
determinant in terms of traces for general $4\times 4$ traceless matrices,%
\begin{equation}
\det \left( \delta _{\alpha }^{\ \beta }+\kappa ~K_{\alpha }^{\ \beta
}\right) =1-\frac{1}{2}~\kappa ^{2}K_{\lambda }^{\ \mu }K_{\mu }^{\ \lambda
}+\frac{1}{3}~\kappa ^{3}K_{\lambda }^{\ \mu }K_{\mu }^{\ \nu }K_{\nu }^{\
\lambda }+\frac{1}{8}~\kappa ^{4}\left( \left( K_{\lambda }^{\ \mu }K_{\mu
}^{\ \lambda }\right) ^{2}-2K_{\lambda }^{\ \mu }K_{\mu }^{\ \nu }K_{\nu
}^{\ \rho }K_{\rho }^{\ \lambda }\right) \ ,
\end{equation}%
an expansion that is valid in any gauge.

The discussion given above, concerning conservation of the energy-momentum
tensor for the massless free field theory, does not apply to the massive
theory that we shall discuss next. \ For the massive free field theory, the
gauge invariance of the model is broken by explicit mass terms. \ In that
case, energy and momentum are still conserved, of course, but a more
detailed proof is required to demonstrate this fact.\newpage

\section*{Properties of the massive\ free field theory}

Consider the massive $T_{\left[ \lambda \mu \right] \nu }$ theory. \ Recall
the free-field Lagrangian density \cite{C1980}%
\begin{eqnarray}
\mathcal{L} &=&-\frac{1}{6}\left( F_{\left[ \lambda \mu \nu \right] \rho
}~F^{\left[ \lambda \mu \nu \right] \rho }-3F_{\left[ \mu \nu \right] }~F^{%
\left[ \mu \nu \right] }\right) +\frac{1}{2}~m^{2}\left( T_{\left[ \lambda
\mu \right] \nu }T^{\left[ \lambda \mu \right] \nu }-2T_{\lambda }T^{\lambda
}\right)  \notag \\
&=&\frac{1}{36}~K_{\mu }^{\ \nu }K_{\nu }^{\ \mu }+\frac{1}{2}~m^{2}\left(
T_{\left[ \lambda \mu \right] \nu }T^{\left[ \lambda \mu \right] \nu
}-2T_{\lambda }T^{\lambda }\right) \ ,  \label{MassiveLagrangian}
\end{eqnarray}%
where the trace of the field is $T_{\lambda }=g^{\mu \nu }T_{\left[ \lambda
\mu \right] \nu }$, and the field strength and its trace are as defined in (%
\ref{FStrength}). \ Once again, it is slightly more compact to write $%
\mathcal{L}$ using $K_{\mu }^{\ \nu }$ as defined in (\ref{KDefn}), to
obtain for the Lorentz metric case%
\begin{equation}
\frac{1}{6}~K_{\alpha \beta }K^{\beta \alpha }=-F_{\left[ \alpha \beta
\gamma \right] \rho }~F^{\left[ \alpha \beta \gamma \right] \rho }+3F_{\left[
\alpha \beta \right] }~F^{\left[ \alpha \beta \right] }\ .
\end{equation}%
The field equations that follow from varying the action for $\mathcal{L}$
are the usual Klein-Gordon equation for $T_{\left[ \lambda \mu \right] \nu }$%
, i.e. the \emph{on-shell condition} 
\begin{equation}
\left( \square +m^{2}\right) T_{\left[ \lambda \mu \right] \nu }=0\ ,
\label{KTFieldEqnRedux}
\end{equation}%
along with the secondary Fierz-Pauli conditions 
\begin{subequations}
\begin{eqnarray}
T_{\ \ \ \ \rho }^{\left[ \mu \rho \right] } &=&0\ ,  \label{OnShell1} \\
\partial _{\lambda }T^{\left[ \lambda \mu \right] \nu } &=&0\ ,
\label{OnShell2} \\
\partial _{\rho }T^{\left[ \mu \nu \right] \rho } &=&0\ .  \label{OnShell3}
\end{eqnarray}%
We shall call these last three equations the \emph{massive half-shell
conditions}, and we designate relations that hold subject to one or more of
these conditions by the \emph{half-shell} symbol \textquotedblleft $\bumpeq $%
\textquotedblright . \ Similarly, we designate relations that hold subject
to the Klein-Gordon relation (\ref{KTFieldEqnRedux}) as well as one or more
of (\ref{OnShell1},\ref{OnShell2},\ref{OnShell3}) by the fully on-shell
(i.e. \emph{full-shell}) symbol \textquotedblleft $\Bumpeq $%
\textquotedblright .

So, as a consequence of (\ref{OnShell1},\ref{OnShell2},\ref{OnShell3}) we
have 
\end{subequations}
\begin{equation}
F_{\left[ \alpha \beta \gamma \right] }^{\ \ \ \ \ \gamma }\bumpeq 0\ ,\ \ \
K_{\mu \nu }\bumpeq K_{\nu \mu }\ ,\ \ \ \partial ^{\mu }K_{\mu \nu }\bumpeq
0\ .  \label{OnShell4}
\end{equation}

\subsection*{Energy-momentum tensors}

Consider 
\begin{equation}
\mathcal{\theta }_{\mu }^{\ \nu }=K_{\mu \alpha }K^{\alpha \nu }-36m^{2}T_{%
\left[ \mu \beta \right] \gamma }T^{\left[ \nu \beta \right] \gamma }-\delta
_{\mu }^{\ \nu }\left( \frac{1}{2}~K_{\alpha \beta }K^{\beta \alpha
}-9m^{2}T_{\left[ \alpha \beta \right] \gamma }T^{\left[ \alpha \beta \right]
\gamma }\right) \ .  \label{KTEnergyMomentum}
\end{equation}%
Note that on-shell, 
\begin{equation*}
\mathcal{\theta }_{\mu \nu }\bumpeq \mathcal{\theta }_{\nu \mu }\ ,\ \ \
\partial ^{\mu }\mathcal{\theta }_{\mu \nu }\Bumpeq 0\ ,
\end{equation*}%
where the latter conservation is established in detail in an Appendix. \
This energy-momentum tensor has the interesting feature that the $m^{2}$
term drops out of the 4D trace.%
\begin{equation}
\mathcal{\theta }_{\mu }^{\ \mu }=-K_{\alpha \beta }K^{\beta \alpha }\ .
\label{KTTrace}
\end{equation}%
Also, by direct calculation, any $\delta _{\mu }^{\ \nu }\left( \cdots
\right) $ terms in $\mathcal{\theta }_{\mu }^{\ \nu }$ do not contribute in
the field equation (\ref{FE}). \ Moreover, any such $\delta _{\mu }^{\ \nu
}\left( \cdots \right) $ terms would contribute nothing to an interaction
written as $K_{\mu \nu }\mathcal{\theta }^{\nu \mu }$ because $K_{\mu \nu }$
is traceless. \ Similarly, any conformal improvement to $\mathcal{\theta }%
_{\mu \nu }$ of the form $\left( \partial _{\mu }\partial _{\nu }-\eta _{\mu
\nu }\square \right) \left( \cdots \right) $ will contribute nothing to
either the field equation (\ref{FE}) or to the \emph{bulk} action obtained
by integrating $K_{\mu \nu }\mathcal{\theta }^{\nu \mu }$ because of the
vanishing trace and divergence kinematic conditions on $K_{\mu \nu }$. \ But
note that, in principle, \emph{boundary} contributions to the action are
possible from coupling to a conformal improvement term of this form
following an obvious integration by parts.

\section*{Interaction Lagrangian to lowest order}

The field equation may be written more compactly as 
\begin{equation}
\left( \square +m^{2}\right) T_{\left[ \lambda \mu \right] \nu }=\kappa
P_{\lambda \mu \nu ,\alpha \beta \gamma }\partial ^{\alpha }\Theta ^{\beta
\gamma }\ ,  \label{FEAgain}
\end{equation}%
where we have defined a symmetrizing tensor%
\begin{equation}
P_{\lambda \mu \nu ,\alpha \beta \gamma }=2\varepsilon _{\lambda \mu \alpha
\beta }\eta _{\nu \gamma }+\varepsilon _{\nu \mu \alpha \beta }\eta
_{\lambda \gamma }-\varepsilon _{\nu \lambda \alpha \beta }\eta _{\mu \gamma
}\ .
\end{equation}%
Note that any $\delta _{\beta }^{\ \gamma }\left( \cdots \right) $ or $%
\partial _{\beta }\left( \cdots \right) ^{\gamma }$ terms in $\Theta _{\beta
}^{\ \gamma }$ will give no contribution to the field equation because $%
\varepsilon ^{\lambda \mu \alpha \beta }T_{\left[ \lambda \mu \right] \beta
}=0$ and $\varepsilon ^{\lambda \mu \alpha \beta }\partial _{\alpha
}\partial _{\beta }\left( \cdots \right) =0$.

Ignoring terms of $O\left( \kappa ^{2}\right) $, to obtain the field
equations to $O\left( \kappa \right) $ the previous massive field
energy-momentum tensor must be augmented by adding a manifestly\ conserved,
symmetric \textquotedblleft improvement\textquotedblright : 
\begin{eqnarray}
\Theta _{\beta }^{\ \gamma } &=&\theta _{\beta }^{\ \gamma }-36\tau _{\beta
}^{\ \gamma }\ ,\ \ \ \partial ^{\beta }\tau _{\beta }^{\ \gamma }\equiv 0\ ,
\\
\tau _{\beta }^{\ \gamma } &\equiv &\square \left( T_{\left[ \beta b\right]
c}T^{\left[ \gamma b\right] c}\right) -\partial _{\beta }\partial ^{a}\left(
T_{\left[ ab\right] c}T^{\left[ \gamma b\right] c}\right) -\partial ^{\gamma
}\partial _{a}\left( T^{\left[ ab\right] c}T_{\left[ \beta b\right]
c}\right) +\delta _{\beta }^{\ \gamma }\partial ^{d}\partial _{a}\left( T^{%
\left[ ab\right] c}T_{\left[ db\right] c}\right) \ .  \notag
\end{eqnarray}

A Lagrangian which gives the field equation to $O\left( \kappa \right) $ is
then obtained by adding to the free field massive Lagrangian (\ref%
{MassiveLagrangian}) $O\left( \kappa \right) $ interactions \emph{suggested}
by the form $K_{\alpha }^{\ \beta }\Theta _{\beta }^{\ \alpha }$, namely,%
\begin{eqnarray}
\mathcal{L}_{int} &=&-\frac{1}{3}~\kappa K_{\alpha }^{\ \beta }K_{\beta }^{\
\gamma }K_{\gamma }^{\ \alpha }-36\kappa T^{\left[ \lambda \mu \right] \nu
}P_{\lambda \mu \nu ,\alpha \beta \gamma }\partial ^{\alpha }\left( \left(
\square +m^{2}\right) \left( T_{\left[ \beta b\right] c}T^{\left[ \gamma b%
\right] c}\right) -\partial ^{\gamma }\partial _{a}\left( T^{\left[ ab\right]
c}T_{\left[ \beta b\right] c}\right) \right)  \notag \\
&=&-\frac{1}{3}~\kappa K_{\alpha }^{\ \beta }K_{\beta }^{\ \gamma }K_{\gamma
}^{\ \alpha }-108\kappa \varepsilon ^{\lambda \mu \alpha \beta }T_{\left[
\lambda \mu \right] \gamma }\partial _{\alpha }\left( \left( \square
+m^{2}\right) \left( T_{\left[ \beta b\right] c}T^{\left[ \gamma b\right]
c}\right) -\partial ^{\gamma }\partial _{a}\left( T^{\left[ ab\right] c}T_{%
\left[ \beta b\right] c}\right) \right) \ .
\end{eqnarray}%
The resulting action due to $\mathcal{L}_{int}$ is of course%
\begin{equation}
\mathcal{A}_{int}=\int \mathcal{L}_{int}d^{4}x\ .
\end{equation}%
So then, by varying $T^{\left[ \lambda \mu \right] \nu }$ in\ $\mathcal{A}%
_{int}$ the contributions to the field equations follow from%
\begin{eqnarray}
\delta \mathcal{A}_{int} &=&-\kappa \int \left( \delta K_{\alpha }^{\ \beta
}\right) K_{\beta }^{\ \gamma }K_{\gamma }^{\ \alpha }d^{4}x \\
&&-108\kappa \int \left( \delta T_{\left[ \lambda \mu \right] \nu }\right)
\varepsilon ^{\lambda \mu \alpha \beta }\partial _{\alpha }\left( \left(
\square +m^{2}\right) \left( T_{\left[ \beta b\right] c}T^{\left[ \nu b%
\right] c}\right) -\partial ^{\nu }\partial _{a}\left( T^{\left[ ab\right]
c}T_{\left[ \beta b\right] c}\right) \right) d^{4}x  \notag \\
&&-108\kappa \int T_{\left[ \lambda \mu \right] \nu }\varepsilon ^{\lambda
\mu \alpha \beta }\partial _{\alpha }\left( \left( \square +m^{2}\right)
\delta \left( T_{\left[ \beta b\right] c}T^{\left[ \nu b\right] c}\right)
-\partial ^{\nu }\partial _{a}~\delta \left( T^{\left[ ab\right] c}T_{\left[
\beta b\right] c}\right) \right) d^{4}x\ .  \notag
\end{eqnarray}%
However, upon integrating by parts the terms in the last line give \emph{no}
contributions to the bulk field equations at $O\left( \kappa \right) $
because of the $O\left( \kappa ^{0}\right) $ on-shell conditions, (\ref%
{OnShell3}) and (\ref{KTFieldEqnRedux}). \ These terms might be important at 
$O\left( \kappa ^{2}\right) $, but they have no effect at $O\left( \kappa
\right) $.

Rewriting the $K$ trilinear variation%
\begin{equation}
-\kappa \int \left( \delta K_{\alpha }^{\ \beta }\right) K_{\beta }^{\
\gamma }K_{\gamma }^{\ \alpha }d^{4}x=-3\kappa \int \left( \partial
_{a}\delta T_{\left[ bc\right] \alpha }\varepsilon ^{abc\beta }\right)
K_{\beta }^{\ \gamma }K_{\gamma }^{\ \alpha }d^{4}x=3\kappa \int \left(
\delta T_{\left[ \lambda \mu \right] \nu }\right) \varepsilon ^{\lambda \mu
\alpha \beta }\partial _{\alpha }\left( K_{\beta }^{\ \gamma }K_{\gamma }^{\
\nu }\right) d^{4}x\ ,
\end{equation}%
the $O\left( \kappa \right) $ variation of the interaction is therefore%
\begin{eqnarray}
\delta \mathcal{A}_{int} &=&3\kappa \int \left( \delta T_{\left[ \lambda \mu %
\right] \nu }\right) \varepsilon ^{\lambda \mu \alpha \beta }\partial
_{\alpha }\left( K_{\beta }^{\ \gamma }K_{\gamma }^{\ \nu }-36\left( \left(
\square +m^{2}\right) \left( T_{\left[ \beta b\right] c}T^{\left[ \nu b%
\right] c}\right) -\partial ^{\nu }\partial _{a}\left( T^{\left[ ab\right]
c}T_{\left[ \beta b\right] c}\right) \right) \right) d^{4}x+O\left( \kappa
^{2}\right)  \notag \\
&=&3\kappa \int \left( \delta T_{\left[ \lambda \mu \right] \nu }\right)
\varepsilon ^{\lambda \mu \alpha \beta }\partial _{\alpha }\Theta _{\beta
}^{\ \nu }d^{4}x+O\left( \kappa ^{2}\right) \ .
\end{eqnarray}%
That is to say,%
\begin{equation}
\delta \mathcal{A}_{int}=\kappa \int \left( \delta T^{\left[ \lambda \mu %
\right] \nu }\right) P_{\lambda \mu \nu ,\alpha \beta \gamma }\partial
^{\alpha }\Theta ^{\beta \gamma }d^{4}x+O\left( \kappa ^{2}\right) \ .
\end{equation}%
This variation thereby gives precisely the RHS of the field equation (\ref%
{FEAgain}) to lowest non-trivial order in $\kappa $.

\section*{Other forms of the field equations and their static solutions}

Given the proposed field equation (\ref{FE}) for $T_{\left[ \lambda \mu %
\right] \nu }$ the on-shell equation for the $T$-field strength $K_{\mu \nu
} $ is%
\begin{equation}
\left( \square +m^{2}\right) K_{\mu \nu }=-18\kappa \square \Theta _{\mu \nu
}+6\kappa \left( \eta _{\mu \nu }\square -\partial _{\mu }\partial _{\nu
}\right) \Theta \ ,  \label{KE}
\end{equation}%
where $\Theta =\Theta _{\lambda }^{\ \lambda }$. \ If $\partial ^{\mu
}\Theta _{\mu \nu }=0$ and $\Theta _{\mu \nu }=\Theta _{\nu \mu }$,\ the RHS
of (\ref{KE}) is conserved, symmetric, and manifestly traceless in 4D, so
the trace and divergences of $K_{\mu \nu }$ are free fields and may be
consistently set to zero. \ Moreover, while $K_{\mu \nu }+K_{\nu \mu }$
couples to $\Theta _{\mu \nu }=\Theta _{\nu \mu }$, the antisymmetric part $%
K_{\mu \nu }-K_{\nu \mu }$ is also a free field and again may be
consistently set to zero.

The field equation (\ref{KE}) is almost familiar. \ Were it not for the
manifestly conserved trace term, $\left( \partial _{\mu }\partial _{\nu
}-\eta _{\mu \nu }\square \right) \Theta $, an obvious inference from (\ref%
{KE}) would be that a more conventional form of massive gravity, such as
that in \cite{FMS1968}, would be related to the on-shell dual theory just by
the identification $K_{\mu \nu }\propto \square h_{\mu \nu }$, where%
\begin{equation}
\left( \square +m^{2}\right) h_{\mu \nu }=\kappa \Theta _{\mu \nu }\ .
\label{CFE}
\end{equation}%
The trace term invalidates this identification, in general. \ But there are
situations where such an identification is essentially correct. \ This is
especially true for static configurations.

In fact, it may be somewhat surprising that static energy-momentum sources 
\emph{do} produce dual fields, given that the source on the RHS of (\ref{FE}%
) is a total divergence. \ This is perhaps more easily seen from (\ref{KE}).
\ Static sources do indeed produce $K_{00}$ fields. \ In that case,%
\begin{equation}
\left( \nabla ^{2}-m^{2}\right) K_{00}=-18\kappa \nabla ^{2}\Theta
_{00}+6\kappa \nabla ^{2}\Theta \ .
\end{equation}%
For either traceless $\Theta _{\mu \nu }$ or stress-free matter, this is
equivalent to the static case of more conventional massive gravity, as given
by (\ref{CFE}), but with $K_{00}\propto \nabla ^{2}h_{00}$.

For example, suppose the energy density is given by a static isotropic
radial electric field around a small ball of charge, with either $H_{00}$ or 
$h_{00}$ due just to the electric field energy, ignoring any fields produced
by the mass density of the ball. \ In that case $\Theta =0$, classically,
and it is not difficult to determine $H_{00}$ and $h_{00}$ fields outside
the charged ball with homogeneous Dirichlet boundary conditions at spatial
infinity. \ The only difference in the functional form of $H_{00}$ and $%
h_{00}$ for these exact solutions is an extra $1/r^{4}$ term in $H_{00}$. \
Explicitly, solving the equations%
\begin{equation}
-\frac{d^{2}}{dr^{2}}H+m^{2}H=\frac{A}{r^{5}}\ ,\ \ \ \ \ -\frac{d^{2}}{%
dr^{2}}h+m^{2}h=\frac{B}{r^{3}}\ ,
\end{equation}%
for $H\left( r\right) =rH_{00}\left( r\right) $ and $h\left( r\right)
=rh_{00}\left( r\right) $, where $A$ and $B$ are constants proportional to
the source's electric charge squared, leads to%
\begin{eqnarray}
H_{00}\left( r\right) &=&C_{1}~\frac{\exp \left( -mr\right) }{r}-\frac{1}{24}%
\frac{Am^{2}}{r^{2}}-\frac{1}{12}\frac{A}{r^{4}} \\
&&+\frac{1}{24}\frac{Am^{3}}{r}\left( \func{Shi}\left( mr\right) \cosh mr-%
\func{Chi}\left( mr\right) \sinh mr\right) \ ,  \notag
\end{eqnarray}%
\begin{eqnarray}
h_{00}\left( r\right) &=&C_{2}~\frac{\exp \left( -mr\right) }{r}-\frac{1}{2}%
\frac{B}{r^{2}} \\
&&+\frac{1}{2}\frac{Bm}{r}\left( \func{Shi}\left( mr\right) \cosh mr-\func{%
Chi}\left( mr\right) \sinh mr\right) \ ,  \notag
\end{eqnarray}%
where the $C$'s are constants of integration determined by boundary
conditions at the surface of the charged ball. \ The same special functions
appear in both cases, namely, 
\begin{equation}
\func{Shi}\left( mr\right) =\int_{0}^{mr}\frac{\sinh \left( t\right) }{t}%
~dt=mr+\frac{1}{18}~m^{3}r^{3}+O\left( r^{5}\right) \ ,
\end{equation}%
\begin{eqnarray}
\func{Chi}\left( mr\right) &=&\func{gamma}+\ln \left( mr\right)
+\int_{0}^{mr}\frac{\cosh \left( t\right) -1}{t}~dt \\
&=&\func{gamma}+\ln \left( mr\right) +\frac{1}{4}~m^{2}r^{2}+\frac{1}{96}%
~m^{4}r^{4}+O\left( r^{5}\right) \ .  \notag
\end{eqnarray}%
A discussion of the phenomenological differences between the two types of
static fields for this example, and for other more realistic source terms,
will be given elsewhere.

\section*{Relation to the Ogievetsky-Polubarinov model}

The analysis of the previous section suggests that a field redefinition may
provide additional insight for the dual theory. \ It does. \ 

Adding and substracting $m^{2}\Theta _{\mu \nu }$\ to the RHS of (\ref{KE})
and moving $\left( \square +m^{2}\right) \Theta _{\mu \nu }$ to the LHS gives%
\begin{equation}
\left( \square +m^{2}\right) H_{\mu \nu }=\kappa \Theta _{\mu \nu }+\frac{%
\kappa }{3m^{2}}\left( \eta _{\mu \nu }\square -\partial _{\mu }\partial
_{\nu }\right) \Theta \ ,  \label{OPE}
\end{equation}%
where the field redefinition is simply 
\begin{equation}
H_{\mu \nu }=\frac{1}{18m^{2}}\left( K_{\mu \nu }+18\kappa \Theta _{\mu \nu
}\right) \ .  \label{HDefn}
\end{equation}%
From the first kinematic constraint in (\ref{Trace&DivConditions}), the
trace $H=H_{\mu }^{\ \mu }$ is then constrained to be%
\begin{equation}
H=\frac{\kappa }{m^{2}}~\Theta \ .  \label{TraceH}
\end{equation}%
This constraint on the trace is consistent with (\ref{OPE}) because, given
that field equation, the difference $H-\kappa \Theta /m^{2}$ is a free
field. \ Similarly, the divergence and antisymmetric parts of the $H$-field
are free and consistently set to zero.

In general, (\ref{HDefn}) is a nonlinear field redefinition, given that $%
\Theta _{\mu \nu }$ depends on the dual field. \ But in the weak-field
limit, outside any non-$T$-field source of energy-momentum, the $H_{\mu \nu
} $ field is just proportional to $K_{\mu \nu }$, hence proportional to the $%
T$-field strength, an expected relation that characterizes free field (or
weak-field) duality.

More importantly, the field equation (\ref{OPE}) is \emph{not} the
conventional one in (\ref{CFE}). \ Rather, (\ref{OPE}) is the field equation
of the Ogievetsky-Polubarinov model for a pure spin 2 massive field \cite{OP}%
. \ This may explain why \cite{CF1980}\ encountered difficulties and could
not obtain a \textquotedblleft perfect\textquotedblright\ dualization
connecting (\ref{FE}) and (\ref{CFE}). \ The latter two equations are \emph{%
not} dual to one another. \ Rather, the interacting massive $T$-theory is
the exact dual of the Ogievetsky-Polubarinov model.

\bigskip

\noindent \textbf{Acknowledgements} \ We thank D. Fairlie and C. Zachos for
comments and suggestions, and the late Peter Freund for \href{https://cgc.physics.miami.edu/PGOFMemories.pdf}%
{many fond memories}. \ This work was supported in part by a University of
Miami Cooper Fellowship.

\bigskip

\section*{Appendix}

On-shell conservation of $\mathcal{\theta }_{\mu }^{\ \nu }$ may be
established as follows.\medskip

\textbf{[Lemma 1]}%
\begin{equation}
\partial ^{\mu }\left( K_{\mu }^{\ \lambda }K_{\lambda }^{\ \nu }-\frac{1}{2}%
\ \delta _{\mu }^{\ \nu }K_{\alpha \beta }K^{\beta \alpha }\right) \bumpeq
3K_{\mu }^{\ \lambda }\varepsilon ^{\alpha \beta \mu \nu }\square T_{\left[
\alpha \beta \right] \lambda }\ .  \tag{A1}  \label{L1}
\end{equation}

Proof:%
\begin{eqnarray*}
\partial ^{\mu }\left( K_{\mu }^{\ \lambda }K_{\lambda }^{\ \nu }\right)
&\bumpeq &K_{\mu }^{\ \lambda }\partial ^{\mu }K_{\lambda }^{\ \nu }=K_{\mu
}^{\ \lambda }\varepsilon ^{\alpha \beta \gamma \nu }\partial ^{\mu }F_{ 
\left[ \alpha \beta \gamma \right] \lambda }\ \ \ \text{using (\ref{OnShell4}%
) and (\ref{KDefn})} \\
&=&K_{\mu }^{\ \lambda }\left( \varepsilon ^{\alpha \beta \gamma \mu
}\partial ^{\nu }+3\varepsilon ^{\alpha \beta \mu \nu }\partial ^{\gamma
}\right) F_{\left[ \alpha \beta \gamma \right] \lambda }\ \ \ \text{\text{%
\href{http://mathworld.wolfram.com/Syzygy.html}{\text{syzygy}}} in 4D \cite%
{Hamermashed}} \\
&\bumpeq &K_{\mu }^{\ \lambda }\partial ^{\nu }K_{\lambda }^{\ \mu }+3K_{\mu
}^{\ \lambda }\varepsilon ^{\alpha \beta \mu \nu }\square T_{\left[ \alpha
\beta \right] \lambda }\ \ \ \text{using (\ref{KDefn}) and (\ref{OnShell2},%
\ref{OnShell3})}
\end{eqnarray*}%
So (\ref{L1}) is established. \ Thus we are led to\medskip

\textbf{[Lemma 2]}%
\begin{equation}
K_{\mu }^{\ \lambda }\varepsilon ^{\alpha \beta \mu \nu }\bumpeq 6F^{\left[
\alpha \beta \nu \right] \lambda }\ .  \tag{A2}  \label{L2}
\end{equation}

Proof:%
\begin{eqnarray*}
K_{\mu }^{\ \lambda }\varepsilon ^{\alpha \beta \mu \nu } &\bumpeq &K_{\ \mu
}^{\lambda }\varepsilon ^{\alpha \beta \mu \nu }=F^{\left[ abc\right]
\lambda }\varepsilon _{abc\mu }\varepsilon ^{\alpha \beta \mu \nu }\ \ \ 
\text{using (\ref{OnShell4}) and (\ref{KDefn})} \\
&=&\delta _{abc}^{\alpha \beta \nu }F^{\left[ abc\right] \lambda }=6F^{\left[
\alpha \beta \nu \right] \lambda }\ \ \ \text{using }\varepsilon _{abc\mu
}\varepsilon ^{\alpha \beta \mu \nu }=-\delta _{abc\mu }^{\alpha \beta \mu
\nu }=+\delta _{abc}^{\alpha \beta \nu }\text{ in 4D}
\end{eqnarray*}%
So (\ref{L2})\ is also established. \ Now, combining (\ref{L1})\ and (\ref%
{L2})\ along with (\ref{KTFieldEqnRedux}) gives immediately\medskip

\textbf{[Lemma 3]}%
\begin{equation}
\partial ^{\mu }\left( K_{\mu }^{\ \lambda }K_{\lambda }^{\ \nu }-\frac{1}{2}%
\ \delta _{\mu }^{\ \nu }K_{\alpha \beta }K^{\beta \alpha }\right) \Bumpeq
-18m^{2}F^{\left[ \alpha \beta \nu \right] \lambda }T_{\left[ \alpha \beta %
\right] \lambda }\ .  \tag{A3}  \label{L3}
\end{equation}%
This leads to a final\medskip

\textbf{[Lemma 4]}%
\begin{equation}
F^{\left[ \alpha \beta \nu \right] \lambda }T_{\left[ \alpha \beta \right]
\lambda }\bumpeq \partial ^{\nu }\left( \frac{1}{2}~T_{\left[ \alpha \beta %
\right] \gamma }T^{\left[ \alpha \beta \right] \gamma }\right) -2\partial
^{\mu }\left( T_{\left[ \mu \beta \right] \gamma }T^{\left[ \nu \beta \right]
\gamma }\right) \ .  \tag{A4}  \label{L4}
\end{equation}

Proof:%
\begin{eqnarray*}
F^{\left[ \alpha \beta \nu \right] \lambda }T_{\left[ \alpha \beta \right]
\lambda } &=&\left( \partial ^{\nu }T^{\left[ \alpha \beta \right] \lambda
}+2\partial ^{\alpha }T^{\left[ \beta \nu \right] \lambda }\right) T_{\left[
\alpha \beta \right] \lambda }\ \ \ \text{definition of }F^{\left[ \alpha
\beta \nu \right] \lambda } \\
&\bumpeq &\partial ^{\nu }\left( \frac{1}{2}~T_{\left[ \alpha \beta \right]
\gamma }T^{\left[ \alpha \beta \right] \gamma }\right) +2\partial ^{\mu
}\left( T_{\left[ \mu \beta \right] \gamma }T^{\left[ \beta \nu \right]
\gamma }\right) \ \ \ \text{using (\ref{OnShell2}) and renaming indices}
\end{eqnarray*}%
So (\ref{L4}) is established. \ Combining (\ref{L3})\ and (\ref{L4})\ we
then obtain%
\begin{equation}
\partial ^{\mu }\left( K_{\mu }^{\ \lambda }K_{\lambda }^{\ \nu }-\frac{1}{2}%
\ \delta _{\mu }^{\ \nu }K_{\alpha \beta }K^{\beta \alpha }\right) \Bumpeq
-18m^{2}\partial ^{\mu }\left( \frac{1}{2}~\delta _{\mu }^{\ \nu }T_{\left[
\alpha \beta \right] \gamma }T^{\left[ \alpha \beta \right] \gamma }-2\left(
T_{\left[ \mu \beta \right] \gamma }T^{\left[ \nu \beta \right] \gamma
}\right) \right) \ .  \tag{A5}
\end{equation}%
That is to say, $\partial ^{\mu }\mathcal{\theta }_{\mu }^{\ \nu }\Bumpeq 0$
with $\mathcal{\theta }_{\mu }^{\ \nu }$ given by (\ref{KTEnergyMomentum}).

\end{document}